\documentclass[aps,amsmath, amssymb, prd,twocolumn,superscriptaddress,floatfix,showpacs,bibliography,10pt]{revtex4-1}

\usepackage{hyperref}
\usepackage[normalem]{ulem} 
\usepackage{tensor}

\usepackage{color}

\begin{document}

\title{Gravitational wave polarization modes in $f(R)$ theories}

\author{H. Rizwana \surname{Kausar}}
\email[]{rizwa\_math@yahoo.com}
\affiliation{Physik-Institut, Universit\"at Z\"urich, Winterthurerstrasse 190, 8057 Z\"urich, Switzerland}
\affiliation{CAMS, UCP Business School, University of Central Punjab, 1-Khayaban-e-Jinnah Road, Johar Town, Lahore, Pakistan}

\author{Lionel \surname{Philippoz}}
\email[]{plionel@physik.uzh.ch}
\affiliation{Physik-Institut, Universit\"at Z\"urich, Winterthurerstrasse 190, 8057 Z\"urich, Switzerland}

\author{Philippe \surname{Jetzer}}
\email[]{jetzer@physik.uzh.ch}
\affiliation{Physik-Institut, Universit\"at Z\"urich, Winterthurerstrasse 190, 8057 Z\"urich, Switzerland}

\date{\today}

\begin{abstract}

Many studies have been carried out in the literature to evaluate the number of polarization modes of gravitational waves in 
modified theories, in particular in $f(R)$ theories. In the latter ones, besides the usual two transverse-traceless tensor modes present in 
general relativity, there are two additional scalar ones: a massive longitudinal mode and a massless transverse mode (the so-called breathing 
mode). This last mode has often been overlooked in the literature, due to the assumption that the application of the Lorenz gauge implies 
transverse-traceless wave solutions. We however show that this is in general not possible and, in particular, that the traceless condition cannot 
be imposed due to the fact that we no longer have a Minkowski background metric. Our findings are in agreement with the results found using the 
Newman-Penrose formalism, and thus clarify the inconsistencies found so far in the literature.
\end{abstract}

\pacs{04.30.-w, 04.50.Kd}

\maketitle

\section{Introduction}\label{section:introduction}

The questions about the concepts of dark matter and dark energy motivated the development of new gravity theories. Most of them are direct 
modifications of general relativity (GR), such as $f(R)$ theories where, in contrast to GR, the Einstein-Hilbert Lagrangian density is replaced by 
a nonlinear function $f(R)$. The nonlinearities lead to different sets of field equations according to the different variational approaches for 
the action \cite{Sotiriou}.

The first variational approach is known as the metric formalism, where fourth-order field equations are derived by varying the action with respect
to the metric tensor $g_{\mu\nu}$. In this case, connections are metric dependent and hence every field in the gravitational sector is coming from 
the metric tensor. The second type of variational approach leads to the Palatini formalism where the metric and the connections are assumed to be 
independent fields, and the action is this time varied with respect to both of them. The field equations remain of second order as is the case of 
the Einstein field equations. A third type is the metric-affine variational approach, which comes if one uses the Palatini variation but also 
includes torsion by assuming nonmetricity of the connections. This last approach is the most general case of $f(R)$ gravity.

The scalar-tensor theories proposed by Brans and Dicke \cite{Brans} with the aim of making the theory of gravity compatible with Mach's principle 
can have a dynamical equivalence to $f(R)$ gravity, the metric formalism of $f(R)$ corresponding to the specific case 
$\omega_\text{BD}=0$ and the Palatini one to $\omega_\text{BD}=-3/2$ (see e.g. \cite{Sotiriou} or \cite{Felice-Tsujikawa}). Scalar-tensor 
theories 
are also of great interest since a coupling between a scalar field and gravity seems to be a generic outcome of the low-energy limit of string 
theories \cite{Casas}. Another interest in the scalar-tensor models lies in the fact that the $f(R)$ theories can be written as the Einstein 
equations plus a scalar field, and thus we could in principle extend the same formalism applied for the scalar-tensor theories to the $f(R)$ field 
equations \cite{Capone}.

The recent detection of gravitational waves (GWs) by the LIGO Collaboration \cite{LIGO1} is a milestone in GWs research and opens 
new perspectives in the study of general relativity and astrophysics. Moreover, future space-borne detectors will offer access to an 
unprecedented signal sensitivity \cite{eLISA}. It is thus worthwhile to explore GWs in alternative theories of gravity, especially in $f(R)$ and 
scalar-tensor theories. The observation of the polarization modes of GWs will be a key tool to obtain valuable information about the astronomical 
objects and physics of the early Universe; depending if additional polarizations are found or not in a detected signal, our knowledge of 
gravitation could have to be extended beyond GR, but we could in any case exclude some theoretical models according to which modes are actually 
detected. Note that even if the recent detection allows one to put contraints on the Compton wavelength of a massive graviton, it 
cannot exclude the existence of non-GR polarization yet, but this will be possible in the future using a network of detectors with different 
orientations \cite{LIGO2}.

Usually, the general procedure to evaluate the power spectrum of the cosmological GWs is to consider the linearized theory by making small metric 
perturbations around the Minkowski metric (e.g. \cite{Maggiore}). The basic idea is to analyze all the physically relevant components of the 
Riemann tensor $R_{\lambda\mu\nu\rho}$, which cause relative acceleration between test particles. In linearized theory, the Riemann tensor can be 
split into six algebraically independent components. Assuming that GWs are propagating in the $z$ direction, the six components can be 
classified into six polarizations modes, namely  $+$, $\times$, $b$, $l$, $x$, and $y$ denoting plus, cross, breathing, longitudinal, vector-$x$, 
and vector-$y$ modes, respectively. According to the rotation symmetry around the propagation axis of the GWs, the $+$ and $\times$ modes can be 
identified with the tensor-type (spin 2) GWs, the $x$ and $y$ modes are vector-type (spin 1) GWs, and the $b$ and $l$ modes are scalar-type (spin 
0) GWs.

In GR, the GWs present only two polarization states, the $+$ and $\times$ modes.  However, when the framework of an alternative theory of gravity 
is considered, the number of non-null components of the Riemann tensor and hence of polarization modes can be greater than two. This is a direct 
consequence of the new field equations which can lead to the existence of additional radiative modes. Using a linearized approach, 
several studies investigated the additional polarization modes in all versions of $f(R)$ theories (\cite{Capo1, Capo2, Capo3, Corda1, Corda2, 
Corda3}, or more recently \cite{Prasia}) and scalar-tensor theories \cite{Capo4}, and it has been shown  that only one massive longitudinal mode 
exists along with the two usual tensorial modes of GR.
In addition to the 
linearized approach, another powerful tool to study the properties of GWs in any metric theory is the Newman-Penrose (NP) approach developed by 
Eardley \emph{et al}. \cite{NP, Eardley}. In their work, they used a null-tetrad basis in order to calculate the NP \cite{Alves1} quantities in 
terms of the irreducible parts of $R_{\lambda\mu\nu\rho}$, namely, the Weyl tensor, the traceless Ricci tensor and the Ricci scalar. They showed 
that six possible modes of GWs polarization can be represented by these non-null NP quantities. Applying this technique, Alves, Miranda and de 
Araujo \cite{Alves1,Alves2} found that scalar-tensor and $f(R)$ theories have, with respect to GR, two additional modes: a longitudinal and a 
breathing mode. Therefore, there seems to be a disagreement between the linearized and the NP approaches.

Adopting the linearized approach, we explore extra polarization modes of GWs in modified $f(R)$ theories. The outline of the paper is as follows: 
in Sec.~\ref{section:lintheory}, we consider the $f(R)$ model with a quadratic term and linearize the field equation for this model in the metric 
formalism. The solutions of GWs arising in this theory are found explicitly and we present how two more polarization modes appear in addition to 
the usual two coming from GR. In Sec.~\ref{section:palatini_metric_affine}, we consider the Palatini formalism for $f(R)$ theories. A summary of 
our results is given in Sec.~\ref{section:conclusion}. In the Appendix, we briefly overview the NP formalism to show the consistency of our 
results.

\section{Polarization Modes in Metric $f(R)$ Theory}\label{section:lintheory}

In this section we closely follow the approach discussed in Ref. \cite{Gair} and rederive the main results leading to the wave equations for GWs.
In a generic $f(R)$ theory, the corresponding action with a generalized function of the Ricci scalar $R$ can be written as
\begin{equation}\label{action1}
S=\frac{1}{2\kappa}\int d^{4}x\sqrt{-g}f(R),
\end{equation}
where $\kappa$ is the coupling constant and $g$ is the determinant of the metric tensor. Varying this action with respect to $g_{\mu\nu}$ yields 
the following set of vacuum field equations for $f(R)$ gravity:
\begin{equation}\label{FES1}
f'(R)R_{\mu\nu}-\frac{1}{2}f(R)g_{\mu\nu}-\nabla_{\mu}
\nabla_{\nu}f'(R)+ g_{\mu\nu} \Box f'(R)=0,
\end{equation}
where $\mu,\nu=0,1,2,3$ and $\Box=\nabla^{\mu}\nabla_{\mu}$ with $\nabla_{\mu}$ being the covariant derivative for $g_{\mu\nu}$. Taking the trace 
of the field equations, we get
\begin{equation}\label{tr1}
f'(R)R +3 \Box f'(R)-2f(R)=0.
\end{equation}
To study gravitational waves, we use the linearized framework as in GR. Considering  the perturbation of the metric from flat Minkowski space such 
that
\begin{eqnarray}\label{g}
g_{\mu\nu}&=&\eta_{\mu\nu}+h_{\mu\nu},
\\\label{R} R&=&R^{(0)}+R^{(1)},
\end{eqnarray}
with $R^{(0)}=0$. The first-order Ricci tensor and Ricci scalar are given, respectively, by
\begin{eqnarray}\label{rt}
R^{(1)}_{\mu\nu}&=&\frac{1}{2}(\partial_\mu\partial_\rho h^\rho_\nu
+\partial_\nu\partial_\rho h^\rho_\mu-\partial_\mu\partial_\nu
h-\Box h_{\mu\nu}), \\\label{rs} R^{(1)}&=&\partial_\mu\partial_\rho
h^{\rho\mu}-\Box h.
\end{eqnarray}
We first discuss the case where $f(R)$ is a polynomial and then the other cases.

\subsection{Polynomial $f(R)$ models}\label{subsection:polynomial}

We shall first consider polynomial $f(R)$ models of the form
\begin{equation}\label{fr1}
f(R)=R+\alpha R^2 + \beta R^3 + \ldots \ .
\end{equation}
For such cases up to first order in $R^{(1)}$, only terms up to $R^2$ contribute to the field equations \eqref{FES1} for which we get
\begin{equation}\label{FES2}
R^{(1)}_{\mu\nu}-\frac{1}{2}R^{(1)}\eta_{\mu\nu}-2\alpha\partial_{\mu}
\partial_{\nu}R^{(1)}+ 2\alpha\eta_{\mu\nu} \Box R^{(1)}=0.
\end{equation}
The trace equation can be written in the form of a Klein-Gordon equation:
\begin{equation}\label{tr2}
\Box R^{(1)} + m^2R^{(1)}= 0,
\end{equation}
where $m^2=-\frac{1}{6\alpha}$. Physically meaningful solutions require $m^2 > 0$ and thus negative values for $\alpha$.\\

In GR, the linearized Einstein field equations can be reduced to the simple wave equation
\begin{equation}
 \Box \bar{h}_{\mu\nu} = - \frac{16 \pi G}{c^4} T_{\mu\nu}
 \end{equation}
if one defines the trace-reversed perturbation
\begin{equation}\label{tracereversed_hbar}
\bar{h}_{\mu\nu} = h_{\mu\nu} - \frac{h}{2} \eta_{\mu\nu}
\end{equation}
or equivalently
\begin{equation}\label{tracereversed_h}
h_{\mu\nu} = \bar{h}_{\mu\nu} - \frac{\bar{h}}{2} \eta_{\mu\nu}
\end{equation}
and then imposes the Lorenz gauge for $\bar{h}_{\mu\nu}$ \cite{Maggiore, Capo5}:
\begin{equation}\label{lg}
\nabla^{\mu}\bar{h}_{\mu\nu}=0.
\end{equation}
We now want to apply this similar standard reasoning within the $f(R)$ framework and find a quantity $\bar{h}_{\mu\nu}$ that satisfies a wave 
equation when linearizing the field equations \eqref{FES1}. It has been shown \cite{Gair} that the appropriate transformation is given by
\begin{equation}\label{h1}
h_{\mu\nu}=\bar{h}_{\mu\nu}-\frac{\bar{h}}{2}\eta_{\mu\nu}-2\alpha
R^{(1)}\eta_{\mu\nu}.
\end{equation}
By taking the trace, we get
\begin{equation}\label{h2}
h=-\bar{h}-8\alpha R^{(1)}.
\end{equation}
We then impose the Lorenz gauge \eqref{lg}, and after inserting equations \eqref{lg} and \eqref{h1} into \eqref{rt} for the Ricci tensor, we 
obtain 
instead
\begin{equation}\label{rt2}
R^{(1)}_{\mu\nu}=\frac{1}{2}\left[4\alpha\partial_\mu\partial_\nu
R^{(1)}-\Box(\bar{h}_{\mu\nu}-\frac{\bar{h}}{2}\eta_{\mu\nu})+2\alpha
\eta_{\mu\nu}\Box R^{(1)}\right].
\end{equation}
Substituting it into Eq.~\eqref{FES2}, we find
\begin{equation}\label{FES3}
-\frac{1}{2}\Box\bar{h}_{\mu\nu}+\frac{1}{4}\eta_{\mu\nu}\Box\bar{h}+3\alpha\eta_{\mu\nu}
\Box R^{(1)}-\frac{1}{2}\eta_{\mu\nu}R^{(1)}=0.
\end{equation}
Using Eq.~\eqref{h1} and the Lorenz gauge given by Eq.~\eqref{lg}, Eq.~\eqref{rs} becomes
\begin{equation}\label{rs2}
R^{(1)}=6\alpha\Box R^{(1)}+\frac{1}{2}\Box \bar{h}
\end{equation}
or
\begin{equation}\label{rs3}
\Box R^{(1)} + m^2 R^{(1)}=  \frac{1}{2} m^2 \Box \bar{h}.
\end{equation}
Comparing this equation with Eq.~\eqref{tr2} it follows that $\Box \bar{h}=0$ has to be fulfilled as well. Inserting $\Box \bar{h}$ as from 
Eq.~\eqref{rs3} into Eq.~\eqref{FES3} we finally get
\begin{equation}\label{gw1}
\frac{1}{2}\Box\bar{h}_{\mu\nu}=0.
\end{equation}
In GR this wave equation is then solved using the Lorenz gauge, which implies transverse wave solutions, and the vanishing of the trace $\bar{h}$. 
The latter quantity being a scalar in GR and thus invariant under coordinate transformations, both $h_{\mu\nu}$ and $\bar{h}_{\mu\nu}$ can be 
traceless at the same time.
However, in $f(R)$ theories, as can be seen from Eq.~\eqref{h2}, by imposing $\bar{h}=0$  one cannot obtain  $h = 0$, since the trace does no 
longer behave as a scalar under coordinate transformations, due to the additional coupled scalar equation for $R^{(1)}$. It has been argued that 
nonetheless it is possible to perform a gauge transformation such that $\bar{h}=0$, in which case the solution of the wave equation \eqref{gw1} 
would thus be the same as in GR, with no additional polarization mode, in particular the breathing mode.
However, we show in the following that, in $f(R)$ theories, one cannot preserve transversality and the traceless condition at the same time. The 
main point being that when considering a gauge transformation in order to get a vanishing $\bar{h}=0$ one has to take into account that the 
background metric is no longer just Minkowski, but due to the fact that $R^{(1)}$ is nonzero, as clearly has to be the case due to
Eq.~\eqref{tr2}, the metric is $\bar{g}_{\mu\nu}$ as induced by $R^{(1)}$.
 
Let us consider a gauge transformation generated by $\xi_{\mu}$, in which case
$\bar{h}_{\mu\nu}$ becomes
\begin{eqnarray}\label{h3_one}
\bar{h}_{\mu\nu}\rightarrow
\bar{h}_{\mu\nu}'=\bar{h}_{\mu\nu}+\xi_{\nu;\mu}+\xi_{\mu;\nu} - \bar{g}_{\mu\nu}\xi_\lambda^{~;\lambda},
\end{eqnarray}
where a semicolon denotes the covariant derivative with respect to the background metric $\bar{g}_{\mu\nu}$, which is thus used for 
raising and lowering indices. Taking the trace of this equation by contracting with $\bar{g}_{\mu\nu}$, we get
\begin{eqnarray}\label{h4}
\bar{h}'=\bar{h}-2\xi_{\lambda}^{~;\lambda}
\end{eqnarray}
Applying $\Box$ on the above equation, we obtain a condition that $\xi_\lambda$ has to fulfill if we require that $\bar{h}'$ is traceless:
\begin{eqnarray}\nonumber
\Box{\bar{h}'}=0~&\Leftrightarrow&~\Box{(\bar{h}-2\xi_{\lambda}^{~;\lambda})}=0,\\\nonumber
&\Leftrightarrow&~\Box{\xi_{\lambda}^{~;\lambda}}=0 \ \Leftrightarrow \ \left(\Box{\xi_{\lambda}}\right)^{;\lambda}=0\\\label{h4_one}
 &\Leftrightarrow&~\Box {\xi_{\lambda}}=0~~\text{or constant}.
\end{eqnarray}
In the second step, we used the fact that $\Box \bar{h}=0$. Now, requiring the Lorenz gauge condition and using the Ricci identity (see e.g. 
\cite{Straumann})
\begin{equation}
\xi^i_{~;l;k} - \xi^i_{~;k;l} = \tensor{R}{^i_{jkl}} \xi^j, 
\end{equation}
we get from the above transformation \eqref{h3_one}
\begin{eqnarray}\label{h5}
\tensor{\bar{h}}{_{\mu}_{\nu}^{;\nu}} \rightarrow
\tensor{\bar{h}{'}}{_{\mu}_{\nu}^{;\nu}} = \tensor{\bar{h}}{_{\mu}_{\nu}^{;\nu}}+\tensor{\xi}{_{\mu}_{;\nu}^{;\nu}}+\bar{R}^{(1)}_{\mu\nu} 
\xi^\nu,
\end{eqnarray}
where  $\bar{R}_{\mu\nu}^{(1)}$ is the Ricci tensor of the background metric. Since $\tensor{\bar{h}}{_{\mu}^{\nu}_{;\nu}}=0$ due to the Lorenz 
gauge 
condition, hence $\tensor{\bar{h}{'}}{_{\mu}^{\nu}_{;\nu}}$ can be zero if and only if $\tensor{\xi}{_{\mu}^{;\nu}_{;\nu}} + 
\bar{R}^{(1)}_{\mu\nu}\xi^\nu=0$; however, one has that $\Box{\xi_{\mu}}=-\bar{R}^{(1)}_{\mu\nu}\xi^\nu \neq 0 \text{ or constant}$. This 
contradicts Eq.~\eqref{h4_one} and tells us that it is in general not possible to achieve both conditions of vanishing trace and transversality at 
the same time.

\subsection{Solutions}\label{subsection:solutions}

Next, we discuss the solutions corresponding to waves in vacuum of the wave equations given by Eq.~\eqref{tr2} and Eq.~\eqref{gw1}, 
respectively. We make the usual plane wave ansatz, corresponding to a standard Fourier decomposition, and we get for Eq.~\eqref{gw1}
\begin{equation}
\bar{h}_{\mu\nu}=\hat{h}_{\mu\nu}(k^\rho)\exp(i k_\rho x^\rho),
\end{equation}
where $k$ is a four vector. We assume that the wave is traveling along the $z$ axis, and due to the Lorenz gauge, from which it follows that the 
wave is transverse with respect to the $z$ axis,
\begin{eqnarray}\label{s3}
k^{\mu}&=&\omega(1,0,0,1),
\end{eqnarray}
where $\omega$ is the angular frequency. Due to the Lorenz gauge condition, we have
\begin{eqnarray}\label{s5}
&&k^{\mu}\hat{h}_{\mu\nu}=0,\\\label{s6}\quad\Rightarrow\quad
&&\hat{h}_{0\nu}+\hat{h}_{3\nu}=0.
\end{eqnarray}
Assuming $\hat{h}_{0\nu}=0$, implies $\hat{h}_{3\nu}=0$. The only nonzero components of $\hat{h}_{\mu\nu}$ are therefore $\hat{h}_{11}$, 
$\hat{h}_{12}$, $\hat{h}_{21}$ and $\hat{h}_{22}$, so in general we can write
\begin{eqnarray}\label{m1}
\hat{h}_{\mu\nu}=\left(
  \begin{array}{cccc}
  0 & 0 & 0 & 0 \\
  0 & \hat{h}^b_{11}+\hat{h}_{11}^+ & \hat{h}_{12}^\times & 0 \\
  0 &  \hat{h}_{21}^\times & \hat{h}_{22}^b-\hat{h}_{22}^+& 0 \\
  0 & 0 & 0 & 0 \\
      \end{array}
\right),
\end{eqnarray}
where the superscript $+$, $\times$ and $b$ denote the plus, the cross and the breathing polarizations modes of the gravitational radiations, 
respectively. Note that we have $\hat{h}^b_{11}=\hat{h}^b_{22}\equiv \hat{h}^b$, $\hat{h}^+_{11}=\hat{h}^+_{22}\equiv \hat{h}^+$, and
$\hat{h}^\times_{12}=\hat{h}^\times_{21}\equiv \hat{h}^\times$. These modes are transverse to the direction of propagation, with the two $+, 
\times$ representing quadrupolar deformations and $b$ representing a monopolar breathing deformation. In GR, the breathing polarization mode 
vanishes due to the traceless condition of $\hat{h}_{\mu\nu}$, and only the two usual $+$ and $\times$ polarizations remain.

The solution of the second wave equation \eqref{tr2} can be written as
\begin{equation}\label{s2}
R^{(1)}=\hat{R}(q^\rho)\exp(i q_\rho x^\rho),
\end{equation}
where $q^\rho$ is a four vector. Considering a wave traveling along the $z$ axis and taking $\Omega$ as its angular frequency, we can write
\begin{equation}\label{s4}
q^{\mu}=(\Omega,0,0,\sqrt{\Omega^2-m^2}).
\end{equation}
Thus Eq.~\eqref{s2} yields the massive scalar longitudinal mode in the $z$ direction. By analogy with our previous notation, the corresponding 
amplitude $\hat{R}$ can be written as
\begin{equation}\label{m4}
 \hat{R} \equiv \hat{h}^l,
\end{equation}
where the superscript $l$ denotes the longitudinal mode. However, the longitudinal mode does not travel with the speed of light c, but instead 
has a group velocity \cite{Capo1,Gair}
\begin{equation}
v(\Omega)= \frac{\sqrt{\Omega^2 - m^2}}{\Omega}~,
\end{equation}
which for $m^2 >0$ leads to $v < c$ and for $\Omega < m$ the wave is decaying.

\subsection{General $f(R)$ model}\label{section:general}

To find the polarization modes for a  general $f(R)$ function, including in particular also $f(R)=\frac{1}{R}$ models, we define a scalar field 
$\phi=f'(R)$ and an effective potential 
\begin{eqnarray}\label{v}
V=f(R)-Rf'(R)
\end{eqnarray}
such that  $f''(R) \neq 0$  and $f'(R)$ is invertible. Then the field equations \eqref{FES1} can be written as
\begin{equation}\label{FE2}
R_{\mu\nu}-\frac{1}{2}Rg_{\mu\nu}=\frac{1}{\phi}\left(\frac{1}{2}g_{\mu\nu}V(\phi)+\nabla_{\mu}
\nabla_{\nu}\phi -g_{\mu\nu} \Box \phi\right).
\end{equation}
Next, we consider small perturbations from Minkowski background equation~\eqref{g} in the metric tensor and in the scalar field $\phi_0$, 
i.e., around the constant scalar curvature, such that
\begin{equation}\label{phi}
\phi=\phi_0+\delta\phi.
\end{equation}
Here, we also assume $\phi_0$ to be a steady minimum for the effective potential $V$, say that is  $V_0$. As the effective scalar field and  the 
effective potential arise directly from the prime derivative $f'(R)$ of the spacetime curvature, hence  the potential presents a square (i.e., 
parabolic) trend, in a function of the effective scalar field, near the minimum $V_0$, i.e., 
\begin{equation}\label{Vmin}
V\simeq V_0+\frac{1}{2}a\delta\phi^2 \Rightarrow \frac{dV}{d\phi}\simeq a\delta \phi,
\end{equation}
where $a$ is a constant (see \cite{Corda4} for more details on this argument). The  linearized field equations up to first order in $h_{\mu\nu}$ 
and 
$\delta \phi$ become \cite{Capo1}
\begin{equation}\label{all1}
R^{(1)}_{\mu\nu}-\frac{1}{2}R^{(1)}\eta_{\mu\nu}=\partial_{\mu}
\partial_{\nu}h_f -\eta_{\mu\nu} \Box h_f,
\end{equation}
where $h_f=\frac{\delta \phi}{\phi_0}$. Modifying Eq.~\eqref{h1} for a general function, $h_f$, we have
\begin{eqnarray}\label{h4_two}
h_{\mu\nu}&=&\bar{h}_{\mu\nu}-\frac{\bar{h}}{2}\eta_{\mu\nu}-h_f\eta_{\mu\nu}.
\end{eqnarray}
Using Eqs.~\eqref{rt} and \eqref{rs} along with the above transformation and imposing the Lorenz gauge condition, the linearized field equations 
become
\begin{equation}\label{gw2}
\Box \bar{h}_{\mu\nu}=0.
\end{equation}
This equation is identical to Eq.~\eqref{gw1}; it can thus be solved in a similar manner and the solution---thus the number of polarizations---is 
identical to the one found for a polynomial model. Hence, we have the $+$, $\times$ and breathing modes in this case, too.
  
For the massive mode, taking the trace of the Eq.~\eqref{FE2},  we have
 \begin{equation}\label{V'}
-R=-\frac{3}{\phi}\Box \phi+\frac{2}{\phi} V(\phi).
\end{equation}
Differentiating Eq.~\eqref{v} with respect to $\phi$, we obtain 
\begin{equation}\label{tr3}
R=-V'(\phi).
\end{equation}
Using this equation in Eq.~\eqref{V'}, we get
 \begin{equation}\label{tr3_bis}
3\Box \phi - 2 V(\phi)+\phi V'(\phi)=0.
\end{equation}
In linearized approximation, applying Eqs.~\eqref{phi} and \eqref{Vmin} and assuming $V_0=0$, we finally get
\begin{equation}\label{all2}
\Box h_f+m^2 h_f=0,
\end{equation}
where $m^2=\frac{a\phi_0}{3}$. The solution to this equation yields the longitudinal mode in the same way as discussed in the previous section.

In conclusion, any metric formulation of $f(R)$ leads to the existence of four polarization modes: two tensor modes and two scalar ones.

\section{Polarization Modes in Palatini formalism}\label{section:palatini_metric_affine}

In this formalism, the action can be defined as \cite{Capo2}
\begin{equation}\label{ma1}
S (g, \Gamma) =\frac{1}{2\kappa}\int d^{4}x\sqrt{-g}f(R)+L_m,
\end{equation}
where $L_m$ is the Lagrangian for the matter field. In this approach, the connection $\Gamma$  is nonmetric (i.e., the connection and the metric 
are considered as independent) but with vanishing torsion. Assuming that the matter Lagrangian does not depend on the dynamical connection, the 
field equations can be achieved by varying the action independently with respect to $g_{\mu\nu}$ and $\Gamma^\lambda_{\mu\nu}$ and are given by
\begin{eqnarray}\label{p2}
f'(R)R_{\mu\nu}-\frac{1}{2}f(R)g_{\mu\nu}=kT_{\mu\nu},\\\label{p3}
\nabla\!_\alpha(\sqrt{-g}f'(R)g^{\mu\nu})=0,
\end{eqnarray}
with $\nabla\!_\alpha$ the covariant derivative with respect to $\Gamma$.

In order to study the propagation of GWs, we first assume the vacuum case, i.e., $T=g^{\mu\nu}T_{\mu\nu}=0$. The trace of Eq.~\eqref{p2} leads 
then to the constraint
\begin{eqnarray}\label{ptr}
f'(R)R-2f(R)=0.
\end{eqnarray}
If $f(R)\neq \alpha R^2$, it can be shown that the associated theory actually corresponds to GR with a cosmological constant \cite{Ferraris}, 
which yields the two usual tensorial polarization modes $+$ and $\times$ \cite{Naef1}.

However, if $f(R)=\alpha R^2$ [which identically satisfies Eq.~\eqref{ptr}], one cannot reduce this function to a usual expansion of the form 
``$R+\text{ corrections}$'' which would be physically meaningful. However, using the approach of \cite{Alves1}, one can nevertheless show that the 
Newman-Penrose quantities $\psi_2$, $\psi_3$ and $\phi_{22}$ vanish, and no additional polarization mode would arise (see the Appendix).

\section{Conclusion}\label{section:conclusion}

In this paper we have discussed GWs polarization modes in $f(R)$ theories. Using the weak-field approximation, we have explicitly shown 
that in the polynomial case of $f(R)$ theories, two scalar modes arise in addition to the two ordinary tensorial modes from standard GR. One of 
these extra modes is a massive longitudinal one, whereas the second one is a massless breathing mode. Extending the discussion to a general 
function $f(R)$, one finds that the field equations are similar to the polynomial case where our argument can be applied once again, thus leading 
to the existence of the two additional scalar modes.

As previously mentioned, one can also check that scalar-tensor theories have the same number of polarization modes since $f(R)$ 
theories are  equivalent under simple transformations \cite{Capone}. However, the mass is in this case $m^2= \frac{a\phi_0}{2 \omega +3}$, 
where $\omega$ is the transforming parameter.

Moreover, in Brans-Dicke theory, one generally finds four modes [especially in the case $\omega_\text{BD}$ which is equivalent to the 
$f(R)$ metric formalism], but should the scalar field be massless ($m^2=0$), then only one scalar mode remains, namely the massless breathing mode 
\cite{Will, Sathyaprakash}. In the specific case $\omega_\text{BD}=-3/2$, the theory however reduces to the Palatini formalism, where the vacuum 
field equations only produce the two usual tensor modes since this formulation is equivalent to GR, up to a cosmological constant.

In Table 1 we summarize our results.
\begin{table}[h]
\caption{\label{tab:table1}
Additional polarizations found in various theories. $+$, $\times$, $b$, $l$ denote the plus, cross, breathing and longitudinal modes, 
respectively.}
\begin{ruledtabular}
\begin{tabular}{lll}
Theories & Polarization modes\\ \hline
Metric $f(R)$ gravity  & $+,~\times,~ b, ~l$\\
Palatini $f(R)$ gravity & $+,~\times$\\
Scalar-tensor theory (massive) & $+,~\times,~ b,~ l$\\
Brans-Dicke theory (massive) & $+,~\times,~ b,~ l$\\
Brans-Dicke theory (massless)& $+,~\times,~ b$\\
\end{tabular}
\end{ruledtabular}
\end{table}

We would like to mention that our work is in agreement with the results obtained by Alves, Miranda et de Araujo \cite{Alves2, Alves1} where they 
use the
Newman-Penrose approach. In the Appendix, we shortly summarize the related formalism. Additionally, GWs polarizations have also 
recently been studied in $f(R,T)$ models---where $T$ is the trace of the energy momentum tensor---again in the NP formalism \cite{Alves3}. Since 
one needs to examine the theory in a region far from the source of GWs, where $T=0$, the number of polarization modes is again the same as in 
$f(R)$ theories, i.e., four modes.

Recent studies have shown that in some extensions of $f(R)$ theories, the polarization content can greatly vary. For instance, in 
the so-called $F(T)$ theories (where $T$ is the torsion scalar in teleparallelism), there is an equivalence with GR, and thus no additional GWs 
modes \cite{Bamba}. On the other hand, a detailed study of the GWs solutions of fourth-order gravity has shown that, in general, besides the two 
usual massless solutions, there are two further massive modes with finite-distance interaction \cite{Stabile}.

Finally, we note that if GWs present nontensorial polarization modes as discussed in this paper, a measured signal, for instance a stochastic 
cosmological background of GWs, would consist of a mixture of all those modes. Through the analysis of such a signal, the existence of scalar 
and/or vector modes could help to discriminate between the possible theories of gravity beyond GR or to set some bounds on the respective 
intensity of each mode if only tensorial polarizations are present.\\

\begin{acknowledgments}
H.R.K was supported by a Swiss Government Excellence Scholarship for the academic year 2014-2015 and L.P and P.J thank the Swiss National 
Foundation for support. We also thank the referee for the useful comments.
\end{acknowledgments}

\appendix*
\section{\uppercase{Newman-Penrose Formalism}}
\subsection{Overview}
In this section, we give a short overview of the Newman-Penrose (NP) formalism to find extra polarization modes; more details can be found in the 
appropriate references \cite{Alves1, Alves2, NP}.

To define the NP quantities which corresponds to the six polarization modes of GWs, one can define a system of four linearly independent vectors 
$(e_t,~ e_x,~ e_y,~ e_z)$ at any point of space, which are called tetrads. From this system one can introduce a particular tetrad, known as the NP 
tetrad, denoted as  $k,~l,~m,~\bar{m}$ \cite{NP}. The first two of these vectors are real null vectors defined as
\begin{equation}\label{n1}
k=\frac{1}{\sqrt{2}}(e_t+e_z), \quad l=\frac{1}{\sqrt{2}}(e_t-e_z),
\end{equation}
whereas the other two null vectors $m$ and $\bar{m}$ are complex conjugates of each other and defined as
\begin{equation}\label{n2}
m=\frac{1}{\sqrt{2}}(e_x+ie_y), \quad \bar{m}=\frac{1}{\sqrt{2}}(e_x-ie_y).
\end{equation}
The so-defined tetrad vectors obey the following relations:
\begin{equation}\label{n3}
-k.l=m.\bar{m}=1, \quad k.m=k.\bar{m}=l.m=l.\bar{m}=0.
\end{equation}
The Riemann tensor $R_{\lambda\mu\nu\rho}$ can be split into irreducible parts: a ten-component Weyl tensor ($\psi$'s), a nine-component 
traceless Ricci tensor ($\phi$'s) and a curvature scalar ($\Lambda$). The total number of independent components can however be reduced to six 
\cite{Eardley} by assuming nearly plane wave and making use of the properties of the Riemann tensor. In order to describe a wave in a generic 
metric theory in a null frame, one can associate those six independent components to the following quantities: $\psi_2$, $\phi_{22}$, $\psi_3$ and 
$\psi_4$. $\psi_3$ and $\psi_4$ are complex, and thus each one corresponds to two independent polarizations. These NP quantities are related to 
the following components of the Riemann tensor in the null tetrad basis:
\begin{eqnarray}
\label{np1} \psi_2=-\frac{1}{6}R_{lklk} &\sim& \text{longitudinal scalar mode},\\
\label{np2} \psi_3=-\frac{1}{2}R_{lkl\bar{m}} &\sim& \text{vector-$x$ and $y$ modes},\\
\label{np3} \psi_4=-R_{l\bar{m}l\bar{m}} &\sim& \text{+,$\times$ tensorial modes},\\
\label{np4} \phi_{22}=-R_{lml\bar{m}} &\sim& \text{breathing scalar mode}.\\
\end{eqnarray}
The following relations for the Ricci tensor and the Ricci scalar hold:
\begin{eqnarray}
\label{np5} R_{lk}&=&R_{lklk},\\
\label{np6} R_{ll}&=& 2R_{lml\bar{m}}, \\
\label{np7} R_{lm}&=&R_{lklm},\\
\label{np8} R_{l\bar{m}}&=&R_{lkl\bar{m}},\\
\label{np9} R&=&-2R_{lk}=-2R_{lklk}.
\end{eqnarray}
The field equations \eqref{FES1} and \eqref{tr1} can be written in the following form:
\begin{widetext}
\begin{eqnarray}\label{np11}
R_{\mu\nu}&=&\frac{1}{f'(R)}\left[\frac{1}{2}f(R)g_{\mu\nu}+\nabla_{\mu} \nabla_{\nu}f'(R)-g_{\mu\nu} \Box f'(R)\right],\\ \nonumber\\
\label{np12} R&=&
\frac{2f(R)-3 \Box f'(R)}{f'(R)}.
\end{eqnarray}
\end{widetext}
In this approach, one first determines the Ricci scalar \eqref{np12} (by considering the trace of the field equation) and then substitutes it into 
Eq.~\eqref{np11}. By writing the coordinates of the Ricci tensor in the NP tetrad, one can finally determine the possible polarization modes with 
the help of the previously described Newton-Penrose quantities. For instance for the model $f(R)=R+\alpha R^2$ studied in 
Sec.~\ref{subsection:polynomial}, Eq.~\eqref{np11} and its solution correspond to our previous Eq.~\eqref{tr2} and Eq.~\eqref{s2}. Hence 
substituting the solution given in Eq.~\eqref{s2} into Eq.~\eqref{np11}, we finally get the following nonzero components for the Ricci tensor 
\cite{Alves1}:
\begin{eqnarray}\label{np13}
R_{tt}&=&\frac{-1}{2}(4\alpha q^2+1)R,\\
R_{tz}&=&2\alpha q \sqrt{q^2+\frac{1}{6\alpha}R},\\
R_{zz}&=&\frac{1}{6}(-12\alpha q^2+1)R.
\end{eqnarray}
Using Eqs.~\eqref{np1}-\eqref{np9}, one finds the following NP quantities:
\begin{eqnarray}\label{np14}
\psi_2\neq0, \quad \psi_3=0, \quad \psi_4\neq0, \quad \phi_{22}\neq0.
\end{eqnarray}
Thus we get four polarization modes for the GWs:  $+$, $\times$, $b$ and $l$.

\subsection{Case $f(R)=\alpha R^2$ in Palatini}

We give here the proof that the case $f(R)=\alpha R^2$ also does not give rise to additional polarization modes. First considering a theory of 
the form $f(R)=\alpha R^{-\beta}$, one finds
\begin{eqnarray}
 R_{\mu\nu} & = & - \frac{1}{2\beta} R g_{\mu\nu},\\
 R & = & -\frac{2}{\beta}R,
\end{eqnarray}
similarly to Eqs.~\eqref{np11} and \eqref{np12}. If $\beta\neq -2$, the theory reduces to GR with $R=0$, $R_{\mu\nu}=0$, and hence two tensor 
polarization modes. Otherwise, the field equation reads
\begin{equation}
 R_{\mu\nu} = \frac{1}{4} R g_{\mu\nu},
\end{equation}
which implies that the following Newman-Penrose parameters must be zero:
\begin{equation}
 \psi_2 = \psi_3 = \phi_{22} = 0.
\end{equation}
This, once again, corresponds to the two tensor modes $+$ and $\times$.

\bibliography{gw_paper_bibliography}

\end{document}